\documentclass[aps,prc,twocolumn,groupedaddress,amsmath,amssymb,floatfix,superscriptaddress]{revtex4}
\usepackage{multirow}
\usepackage{graphicx}
\usepackage{color}
\usepackage{bm}
\usepackage{comment}
\usepackage{tablefootnote}
\usepackage{threeparttable}
\usepackage{hyperref}

\newcommand{\PerTonDay}{(ton\,day)$^{-1}$}
\newcommand{\PerTonYr}{(ton\,yr)$^{-1}$}

\begin{document}

\bibliographystyle{h-physrev3}

\title{Search for the Majorana Nature of Neutrinos in the \\ 
Inverted Mass Ordering Region with KamLAND-Zen}

\newcommand{\tohoku}{\affiliation{Research Center for Neutrino
    Science, Tohoku University, Sendai 980-8578, Japan}}
\newcommand{\ipmu}{\affiliation{Kavli Institute for the Physics and Mathematics of the Universe (WPI), 
    The University of Tokyo Institutes for Advanced Study, 
    The University of Tokyo, Kashiwa, Chiba 277-8583, Japan}}
\newcommand{\kyoto}{\affiliation{Kyoto University, Department of Physics, 
    Kyoto 606-8502, Japan}}
\newcommand{\osaka}{\affiliation{Graduate School of 
    Science, Osaka University, Toyonaka, Osaka 560-0043, Japan}}
\newcommand{\rcnp}{\affiliation{Research Center for Nuclear Physics, 
    Osaka University, Ibaraki, Osaka 567-0047, Japan}}
\newcommand{\tokushima}{\affiliation{Department of Physics, 
    Tokushima University, Tokushima 770-8506, Japan}}
\newcommand{\tokushimags}{\affiliation{Graduate School of Integrated Arts and Sciences, 
    Tokushima University, Tokushima 770-8502, Japan}}
\newcommand{\lbl}{\affiliation{Nuclear Science Division, Lawrence Berkeley National Laboratory,
    Berkeley, California 94720, USA}}
\newcommand{\hawaii}{\affiliation{Department of Physics and Astronomy,
    University of Hawaii at Manoa, Honolulu, Hawaii 96822, USA}}
\newcommand{\mitech}{\affiliation{Massachusetts Institute of Technology, 
    Cambridge, Massachusetts 02139, USA}}
\newcommand{\ut}{\affiliation{Department of Physics and
    Astronomy, University of Tennessee, Knoxville, Tennessee 37996, USA}}
\newcommand{\tunl}{\affiliation{Triangle Universities Nuclear Laboratory, Durham, 
    North Carolina 27708, USA; \\
    Physics Departments at Duke University, Durham, North Carolina 27708, USA; \\
    North Carolina Central University, Durham, North Carolina 27707, USA; \\
    and The University of North Carolina at Chapel Hill, Chapel Hill, North Carolina 27599, USA}}
\newcommand{\vt}{\affiliation{Center for Neutrino
   Physics, Virginia Polytechnic Institute and State University, Blacksburg,
   Virginia 24061, USA}}
\newcommand{\washington}{\affiliation{Center for Experimental Nuclear Physics and Astrophysics, 
    University of Washington, Seattle, Washington 98195, USA}}
\newcommand{\nikhef}{\affiliation{Nikhef and the University of Amsterdam, 
    Science Park, Amsterdam, the Netherlands}}
\newcommand{\tokyo}{\affiliation{Center for Nuclear Study, The University of Tokyo, 
    Tokyo 113-0033, Japan}}
\newcommand{\gppu}{\affiliation{Graduate Program on Physics for the Universe, Tohoku University, Sendai 980-8578, Japan}}
\newcommand{\frontier}{\affiliation{Frontier Research Institute for Interdisciplinary Sciences, Tohoku University, Sendai, 980-8578, Japan}}
\newcommand{\bu}{\affiliation{Boston University, Boston, Massachusetts 02215, USA}}
\newcommand{\chapel}{\affiliation{UNC Physics and Astronomy, 120 E. Cameron Ave., Phillips Hall CB3255, Chapel Hill, NC 27599}}
\newcommand{\tohokuphys}{\affiliation{Department of Physics, Tohoku University, Sendai, 980-8578, Japan}}

\newcommand{\atimperialnow}{\altaffiliation
    {Present address: Imperial College London, Department of Physics, 
    Blackett Laboratory, London SW7 2AZ, UK}}
\newcommand{\aticrrnow}{\altaffiliation
    {Present address: Kamioka Observatory, Institute for Cosmic-Ray Research, 
    The University of Tokyo, Hida, Gifu 506-1205, Japan}}
\newcommand{\atqstnow}{\altaffiliation
    {Present address: National Institutes for Quantum and Radiological Science 
    and Technology (QST), Sendai 980-8579, Japan}}
\newcommand{\atmephinow}{\altaffiliation
    {Present address: National Research Nuclear University ``MEPhI'' (Moscow Engineering Physics Institute), 
    Moscow, 115409, Russia}}
\newcommand{\atalabamanow}{\altaffiliation
    {Present address: Department of Physics and Astronomy, University of Alabama, 
    Tuscaloosa, AL 35487, USA}}

%
%
\author{S.~Abe}\tohoku
\author{S.~Asami}\tohoku
\author{M.~Eizuka}\tohoku
\author{S.~Futagi}\tohoku
\author{A.~Gando}\tohoku
\author{Y.~Gando}\tohoku
\author{T.~Gima}\tohoku
\author{A.~Goto}\tohoku
\author{T.~Hachiya}\tohoku
\author{K.~Hata}\tohoku
\author{S.~Hayashida}\atimperialnow\tohoku
\author{K.~Hosokawa}\aticrrnow\tohoku
\author{K.~Ichimura}\tohoku
\author{S.~Ieki}\tohoku
\author{H.~Ikeda}\tohoku
\author{K.~Inoue}\tohoku\ipmu
\author{K.~Ishidoshiro}\tohoku
\author{Y.~Kamei}\tohoku
\author{N.~Kawada}\tohoku
\author{Y.~Kishimoto}\tohoku\ipmu
\author{M.~Koga}\tohoku\ipmu
\author{M.~Kurasawa}\tohoku
\author{N.~Maemura}\tohoku
\author{T.~Mitsui}\tohoku
\author{H.~Miyake}\tohoku
\author{T.~Nakahata}\tohoku
\author{K.~Nakamura}\tohoku
\author{K.~Nakamura}\tohoku
\author{R.~Nakamura}\tohoku
\author{H.~Ozaki}\tohoku\gppu
\author{T.~Sakai}\tohoku
\author{H.~Sambonsugi}\tohoku
\author{I.~Shimizu}\tohoku
\author{J.~Shirai}\tohoku
\author{K.~Shiraishi}\tohoku
\author{A.~Suzuki}\tohoku
\author{Y.~Suzuki}\tohoku
\author{A.~Takeuchi}\tohoku
\author{K.~Tamae}\tohoku
\author{K.~Ueshima}\atqstnow\tohoku
\author{H.~Watanabe}\tohoku
\author{Y.~Yoshida}\tohoku
\author{S.~Obara}\atqstnow\frontier
\author{A.K.~Ichikawa}\tohokuphys

\author{D.~Chernyak}\atalabamanow\ipmu
\author{A.~Kozlov}\atmephinow\ipmu

\author{K.Z.~Nakamura}\kyoto

\author{S.~Yoshida}\osaka

\author{Y.~Takemoto}\aticrrnow\rcnp
\author{S.~Umehara}\rcnp

\author{K.~Fushimi}\tokushima
\author{K.~Kotera}\tokushimags
\author{Y.~Urano}\tokushimags

\author{B.E.~Berger}\ipmu\lbl
\author{B.K.~Fujikawa}\ipmu\lbl

\author{J.G.~Learned}\hawaii
\author{J.~Maricic}\hawaii

\author{S.N.~Axani}\mitech
\author{J.~Smolsky}\mitech
\author{Z.~Fu}\mitech
\author{L.A.~Winslow}\mitech

\author{Y.~Efremenko}\ipmu\ut

\author{H.J.~Karwowski}\tunl
\author{D.M.~Markoff}\tunl
\author{W.~Tornow}\ipmu\tunl

\author{S.~Dell'Oro}\vt
\author{T.~O'Donnell}\vt

\author{J.A.~Detwiler}\ipmu\washington
\author{S.~Enomoto}\ipmu\washington

\author{M.P.~Decowski}\ipmu\nikhef

\author{C.~Grant}\bu
\author{A.~Li}\bu\tunl
\author{H.~Song}\bu

\collaboration{KamLAND-Zen Collaboration}\noaffiliation

\date{\today}

\begin{abstract}
The KamLAND-Zen experiment has provided stringent constraints on the neutrinoless double-beta ($0\nu\beta\beta$) decay half-life in $^{136}$Xe using a xenon-loaded liquid scintillator. We report an improved search using an upgraded detector with almost double the amount of xenon and an ultralow radioactivity container, corresponding to an exposure of $970$\,kg\,yr of $^{136}$Xe. These new data provide valuable insight into backgrounds, especially from cosmic muon spallation of xenon, and have required the use of novel background rejection techniques. We obtain a lower limit for the $0\nu\beta\beta$ decay half-life of $T_{1/2}^{0\nu} > 2.3 \times 10^{26}$\,yr at 90\% C.L., corresponding to upper limits on the effective Majorana neutrino mass of 36--156\,meV using commonly adopted nuclear matrix element calculations.

\end{abstract}

\maketitle

The search for neutrinoless double-beta ($0\nu\beta\beta$) decay is the most practical way to probe the Majorana nature of neutrinos. In the context of light Majorana neutrino exchange between two nucleons, the decay rate is proportional to the square of the effective Majorana neutrino mass $\left<m_{\beta\beta}\right> \equiv \left| \Sigma_{i} U_{ei}^{2}m_{\nu_{i}} \right|$, providing information on the absolute neutrino mass scale and mass eigenstate ordering. To date, KamLAND-Zen has provided the most stringent constraint on $\left<m_{\beta\beta}\right>$ of $\gtrsim$\,100\,meV in the quasidegenerate neutrino mass region~\cite{Gando2016}. An improved search probing $\left<m_{\beta\beta}\right>$ below 50\,meV would provide a first test of the Majorana nature of neutrinos in the inverted mass ordering (IO) region  beyond the quasidegenerate mass region. Such searches also test theoretical models predicting $\left<m_{\beta\beta}\right>$ in this range~\cite{Harigaya2012,Asaka2020,Asai2020}.

\mbox{KamLAND-Zen}~\cite{Gando2012a,Gando2012b,Gando2013a,Asakura2016,Gando2016,Gando2019} is a double-beta decay experiment that exploits the existing KamLAND neutrino detector. The $\beta\beta$ decay source is a Xe-loaded liquid scintillator (\mbox{Xe-LS}) contained in a spherical inner balloon (IB) at the center of the detector. The IB is surrounded by 1\,kton of LS (Outer LS) contained in a 13-m-diameter spherical outer balloon made of 135-$\mu$m-thick nylon/EVOH composite film. To detect scintillation light, 1,325 17-inch and 554 20-inch photomultiplier tubes (PMTs) are mounted on the inner surface of the stainless-steel containment tank (SST), providing 34\% solid-angle coverage. The SST is surrounded by a \mbox{3.2~kton} water-Cherenkov outer detector.

The previous search in \mbox{KamLAND-Zen} used 381\,kg of enriched xenon (referred to as \mbox{KamLAND-Zen 400}) and probed $0\nu\beta\beta$ just above the IO region~\cite{Gando2016}. To further improve this limit, the \mbox{KamLAND-Zen} collaboration upgraded the experiment to 745\,kg of enriched xenon (referred to as KamLAND-Zen 800), nearly twice the target mass of the previous experiment. To hold the additional xenon, a larger and cleaner 3.80-m-diameter IB was constructed with better mitigation measures to avoid dust attachment to the balloon surface~\cite{Gando2021}. The \mbox{Xe-LS} consists of 82\% decane and 18\% pseudocumene (1,2,4-trimethylbenzene) by volume, 2.4\,g/liter of the fluor PPO (2,5-diphenyloxazole), and $(3.13 \pm 0.01)$\% by weight of enriched xenon gas. The isotopic abundances in the enriched xenon were measured by a residual gas analyzer to be $(90.85 \pm 0.13)\%$ \mbox{$^{136}$Xe}, $(8.82 \pm 0.01)\%$ \mbox{$^{134}$Xe}.

Science data-taking started on January 2, 2019. The initial data contained $^{222}$Rn~($\tau = 5.5$\,day), introduced by radon emanation from storage tanks and pipelines during \mbox{Xe-LS} filling, and was used for detector calibration. Event positions and energies are reconstructed based on the timing and charge distributions of photoelectrons recorded by the PMTs. The detector Monte Carlo (MC) simulation is based on \texttt{GEANT4}~\cite{Agostinelli2003,Allison2006} and is tuned to reproduce the timing and charge distributions observed in the data. The optical parameters related to the position dependence of the light yield are corrected based on monochromatic $^{214}$Po $\alpha$ decays in the $^{222}$Rn decay chain. The parameters of the detector energy nonlinear response model describing effects from scintillator quenching and Cherenkov light production are constrained to reproduce the measured spectral shape of $^{214}$Bi $\beta + \gamma$ decays. The estimated energy and vertex resolutions in the \mbox{Xe-LS} are 6.7\%/$\sqrt{E({\rm MeV})}$ and 13.7\,cm/$\sqrt{E({\rm MeV})}$, respectively. Using the 2.225\,MeV $\gamma$ rays from the muon-induced neutron captures on protons, the position- and time-dependent fluctuations of the absolute energy scale in the \mbox{Xe-LS} are determined to be less than 1\%. We considered this deviation as a systematic uncertainty on the energy scale conservatively, while its impact is mitigated by the spectral fit for $2\nu\beta\beta$, as discussed later. The outer LS is 1.1 times brighter than the \mbox{Xe-LS}. The tuned MC reproduces the observed vertex distances between sequential $^{214}$Bi-$^{214}$Po decays ($\tau = 237\,\mu {\rm s}$) coming from $^{222}$Rn in the \mbox{Xe-LS}; the average distances are 38.0\,cm and 38.1\,cm with the data and MC, respectively. It also reproduces the energy spectrum of $^{214}$Bi decays including the high energy tail, indicating that the background contribution from energy reconstruction failure in $2\nu\beta\beta$ decays is negligible.

\begin{figure}[t]
\begin{center}
\vspace{-0.5cm}
\includegraphics[width=1.08\columnwidth]{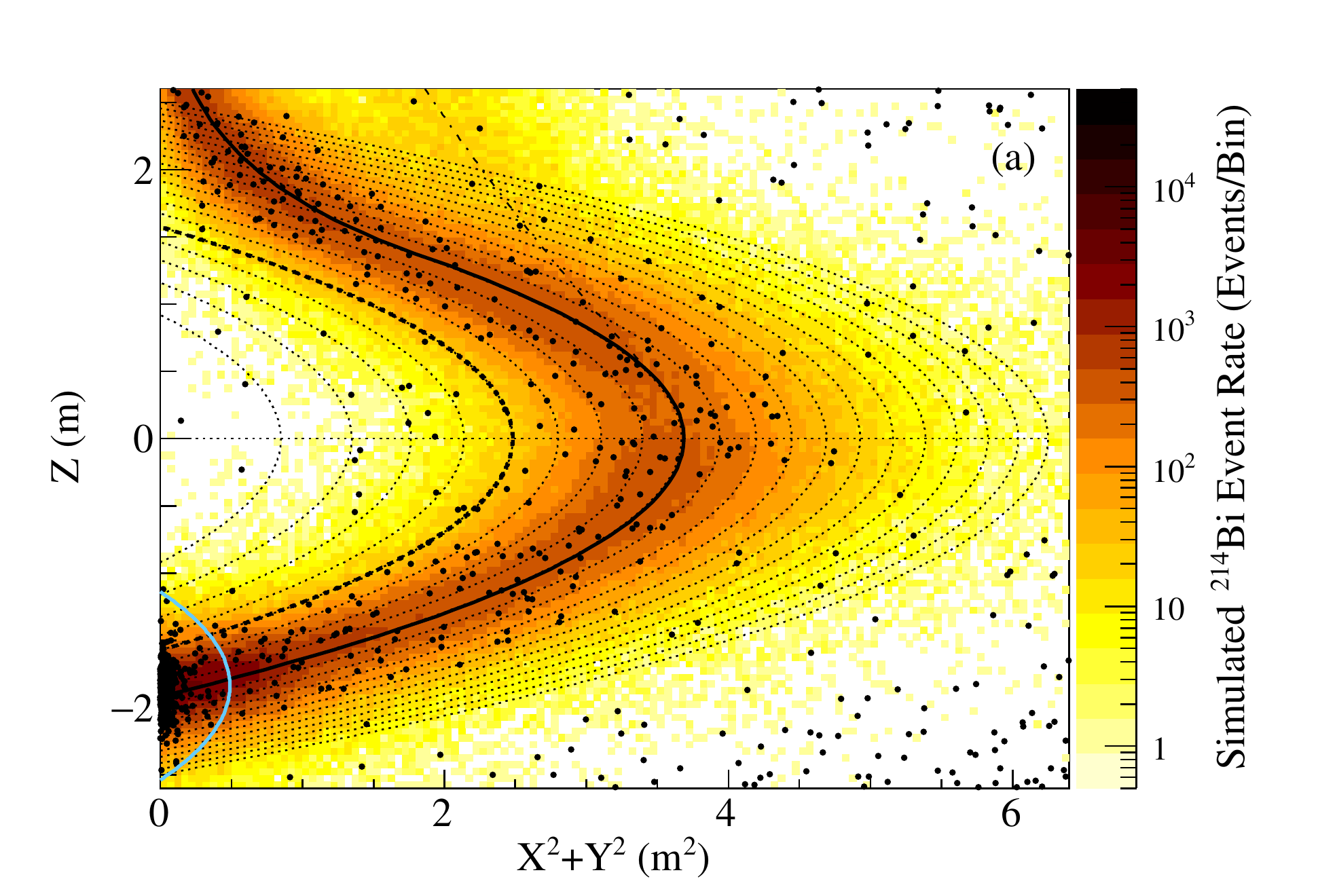}
\\
\vspace{0.1cm}
\includegraphics[width=1.05\columnwidth]{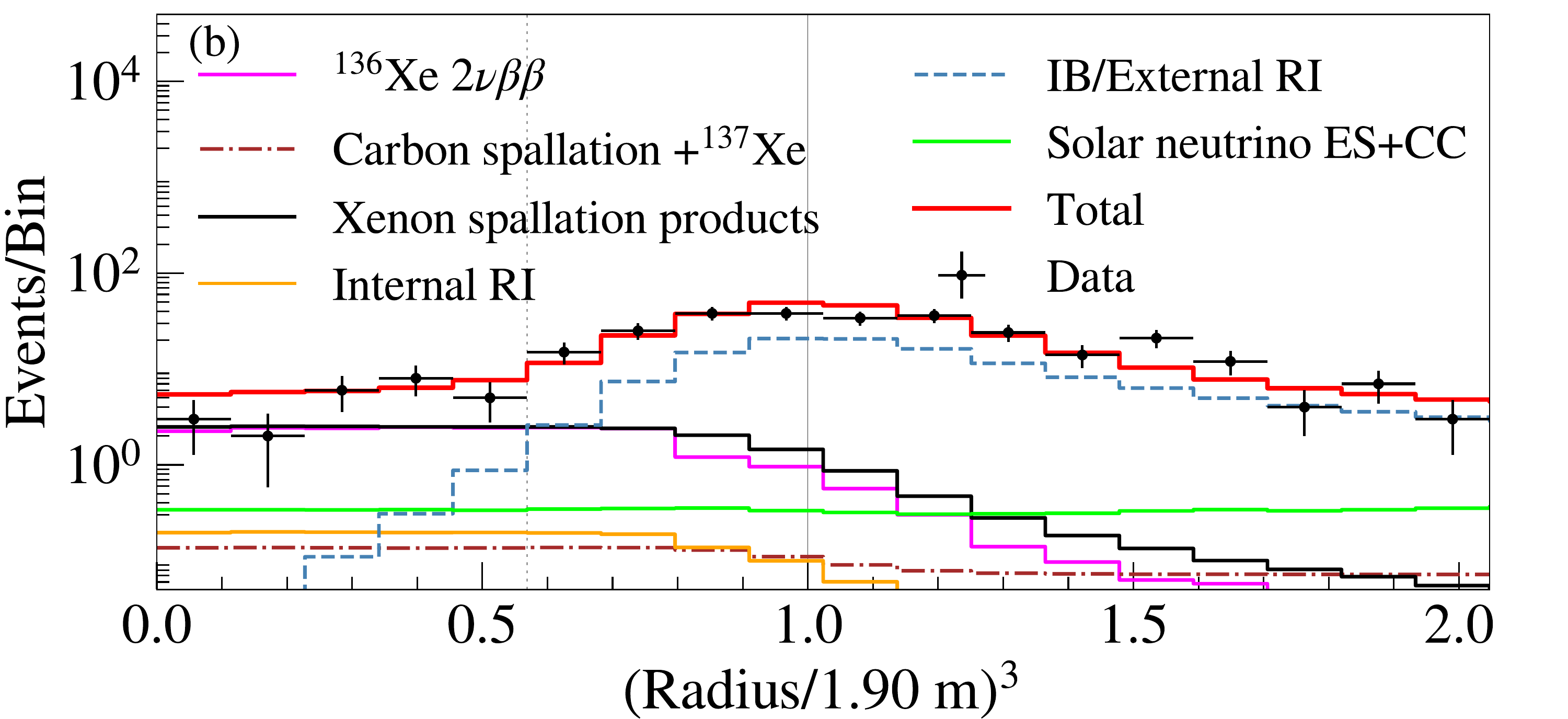}
\vspace{-0.8cm}
\end{center}
\caption{(a) Vertex distribution of candidate SD events (black points) overlaid on $^{214}$Bi background events from the MC simulation (color histogram) in the energy region $2.35 < E < 2.70\,{\rm MeV}$ ($0\nu\beta\beta$ window), with arbitrary normalization. The solid and thick dashed lines indicate the shape of the IB and the 1.57-m-radius spherical volume, respectively. The dot-dashed line indicates the nylon belt suspending the IB. The thin dashed lines illustrate the shape of the equal-volume spherical half-shells, which compose the 2.5-m-radius spherical fiducial volume. The high-count region at the IB bottom indicates the hot spot and is vetoed. (b) $R^{3}$ vertex distribution of candidate SD events in the $0\nu\beta\beta$ window. The curves show the best-fit background model components.}
\vspace{-0.3cm}
\label{figure:vertex}
\end{figure}

We use data collected between February 5, 2019 and May 8, 2021. Candidate events are selected by performing the following series of first-level cuts: (i) the events must be reconstructed within 2.5\,m of the detector center and 0.7\,m away from the bottom hot spot on the IB, which is outlined in Fig.~\ref{figure:vertex}(a). (ii) Muons and events within 2\,ms after muons are rejected. (iii) Sequential radioactive decays are eliminated by a delayed coincidence tag, requiring time and distance between the prompt and delayed events to be less than 1.9\,ms and 1.7\,m, respectively, and a double pulse identification inside a single event acquisition window. Those cuts remove $(99.89 \pm 0.03)$\% of $^{214}$Bi-$^{214}$Po events, and $(97.7 \pm 0.5)$\% of $^{212}$Bi-$^{212}$Po. (iv) Reactor $\overline{\nu}_{e}$ interactions identified by delayed coincidence are rejected. (v) Poorly reconstructed events are rejected to suppress electronic noise and accidental pileup. Real singles events are produced by isotropic scintillation from a single site. In this case, there are correlations between the vertex-to-PMT distance and photon travel time, and between the distance and charge, which can be approximated by simple functions. We define a discriminator based on $\chi^{2}$ tests to assess the agreement with those approximate functions, and identify tag events with high discriminator values as poorly reconstructed events. The overall selection inefficiency is less than 0.1\%.

Background sources for the $0\nu\beta\beta$ search are divided into four categories: (i) radioactive impurities (RI) in the \mbox{Xe-LS}; (ii) external to the \mbox{Xe-LS}, mainly from the IB material; (iii) neutrino interactions; and (iv) cosmogenic spallation products. The inferred contamination of $^{238}$U and $^{232}$Th in the \mbox{Xe-LS} is $(1.5 \pm 0.4)\times 10^{-17}$\,g/g and $(3.0 \pm 0.4)\times  10^{-16}$\,g/g, respectively, based on delayed coincidence measurements of $^{214}$Bi-$^{214}$Po and $^{212}$Bi-$^{212}$Po decays. Those reference calculations for $^{238}$U and $^{232}$Th assume secular equilibrium for comparison with the previously reported values, and are not used for background estimations. We did not find the background peak from $^{110m}$Ag $\beta^{-}$ decays (\mbox{$\tau=360$\,day}, \mbox{$Q = 3.01$\,MeV}) observed previously in \mbox{KamLAND-Zen 400}, caused by contamination from Fukushima-I fallout~\cite{Gando2013a}. We conclude that $^{110m}$Ag was significantly reduced due to radioactive decay, continued purification, and cleaner fabrication of the IB. The primary background sources external to the \mbox{Xe-LS} are $^{238}$U and $^{232}$Th in the IB. The contamination levels of $^{238}$U and $^{232}$Th are $(3 \pm 1)\times 10^{-12}$\,g/g and $(3.8 \pm 0.2)\times 10^{-11}$\,g/g, respectively, and are roughly a factor of 10 smaller compared to those measured on the previous IB~\cite{Gando2016}. The backgrounds from the outer LS and surrounding detector materials are negligibly small.

In the later period of the dataset, we found an increase in the background rate at the IB bottom, possibly due to the settling of dust particles containing radioactive impurities. We performed a search for clusters of $\gamma$-like events using a newly developed spatiotemporal deep neural network model, referred to as KamNet~\cite{Li2023}, which is capable of resolving $\gamma$ cascades from the resulting nonisotropic event topology due to multisite energy deposits. The search identified such an event cluster in the inner volume of the \mbox{Xe-LS} ($R < 1.57$\,m) that is inconsistent with the average background rate, with a \mbox{$p$ value} of only 0.06\% (including the trials penalty). Considering the limited likelihood of contamination during the construction phase, credible sources of time-varying backgrounds are $^{60}$Co in stainless-steel shavings or $^{214}$Bi in thick dust. To avoid this ambiguity in our background modeling, we remove this high-background period from the dataset. The boundaries of the period were defined from the times of the first and last events in the cluster, with an added period on either side corresponding to the average interevent time in the cluster, corresponding to a total of 30.4 days. However, we also performed the analysis with the high-background period included and report the impact on the final result, as discussed later.

Solar neutrinos are an intrinsic background source for the $0\nu\beta\beta$ search. The oscillated neutrino flux can be calculated based on the standard solar model prediction~\cite{Serenelli2011} with three-flavor mixing. The contribution from the elastic scattering (ES) of $^{8}$B solar neutrinos on electrons is estimated to be $(4.9 \pm 0.2) \times 10^{-3} $\,\PerTonDay~in the \mbox{Xe-LS}. In addition, $^{136}$Xe captures $\nu_{e}$ through charged-current (CC) interactions, primarily from $^{7}$Be solar neutrinos, producing $e^{-}$, $^{136}$Cs, and $\gamma$'s from the excited states. The subsequent $\beta + \gamma$ decays of $^{136}$Cs (\mbox{$\tau=19.0$\,day}, \mbox{$Q = 2.548$\,MeV}) produce a background peak around 2.0\,MeV in visible energy, mostly overlapping with  the resolution tail of $2\nu\beta\beta$ decays. The interaction rate is expected to be $(0.8 \pm 0.1)\times  10^{-3}$\,\PerTonDay~based on the cross section calculated in Ref.~\cite{Ejiri2014,Frekers2013}.

Radioactive isotopes produced through cosmic muon spallation of carbon and xenon represent the dominant backgrounds in this analysis. To suppress the spallation backgrounds, second-level cuts are performed with the discrimination parameters based on time intervals ($\Delta T$) from preceding muons, space correlations with vertices of neutron capture $\gamma$ rays (neutron vertices) induced by those muons, and reconstructed muon tracks and shower profiles~\cite{Gando2016,Abe2010,Li2014,Li2015a,Li2015b,Zhang2016}. We apply the following four rejection criteria: (a) events within 150\,ms after muons passing through the LS are rejected. This cut removes 99.4\% of $^{12}$B~($\tau = 29.1\,{\rm ms}$, $Q = 13.4\,{\rm MeV}$). (b) To reduce short-lived carbon spallation backgrounds, mainly from $^{10}$C~($\tau = 27.8\,{\rm s}$, $Q = 3.65\,{\rm MeV}$) and $^{6}$He~($\tau = 1.16\,{\rm s}$, $Q = 3.51\,{\rm MeV}$), we remove events reconstructed within 1.6\,m of neutron vertices with $\Delta T < 180\,{\rm s}$. (c) Events likely to be spallation backgrounds, which have space and time correlations with the preceding muon-induced showers, are rejected. The muon track reconstructed from the timing of the first-arriving photons at the PMTs provides the transverse distance between muons and spallation backgrounds ($l_{\rm trans}$). We apply a newly developed muon shower reconstruction to calculate light intensity profiles along the muon track based on the timing of all the photons. The light profile ($Q_{\rm shower}$) is represented by a function of the longitudinal distance between muon entry and spallation backgrounds ($l_{\rm long}$). We performed the cuts with a new likelihood discriminator, $L_{1}= f_{\rm spall} / f_{\rm acc}$. Here $f_{\rm spall}$ and $f_{\rm acc}$ are the probability density functions (PDFs) for muon-spallation pairs and accidental pairs, respectively; both PDFs are functions of the three parameters: $\Delta T$, $l_{\rm trans}$, and $Q_{\rm shower}$. The cut value of $L_{1}$ was optimized based on the PDFs created from $^{12}$B. The overall rejection efficiencies for $^{10}$C and $^{6}$He by cuts (a)--(c) are $>$99.3\% and $(97.6 \pm 1.7)\%$, respectively, including the uncertainties from the isotope dependence of the PDFs. (d) To reduce the $^{137}$Xe ($\tau = 5.5\,{\rm min}$, $Q = 4.17\,{\rm MeV}$) background, we remove events reconstructed within 1.6\,m of the vertices identified as neutron captures on $^{136}$Xe producing high energy $\gamma$'s ($Q = 4.03\,{\rm MeV}$) with $\Delta T < 27\,{\rm min}$. This cut removes $(74 \pm 7)\%$ of $^{137}$Xe. The dead time introduced by the cuts (a)--(d) is $(14.6 \pm 0.1)\%$. 

The second-level cuts effectively reject carbon spallation backgrounds, however, most xenon spallation backgrounds remain after the cuts because their lifetimes are long (typically several hours). Xenon spallation can be characterized by detecting multiple neutrons emitted via the nucleon evaporation process in neutron-rich isotopes ($^{136}$Xe primarily). To tag the long-lived products effectively, we define another likelihood discriminator, $L_{2} = f_{\rm ll} / f_{\rm acc}$. Here $f_{\rm ll}$ and $f_{\rm acc}$ are the PDFs for long-lived muon-spallation pairs and accidental pairs, respectively, constructed from three parameters: neutron multiplicity, distance to neutron vertices, and $\Delta T$. The cut value on the $L_{2}$ parameter is optimized using an MC simulation discussed later. Events that are not classified as coming from long-lived backgrounds are referred to as ``singles data'' (SD), and the others are referred to as ``long-lived data'' (LD). The total live time for SD and LD is $523.4$\,days and $49.3$\,days, respectively. The exposure of SD, sensitive to $0\nu\beta\beta$ signal, is $970$\,kg\,yr of $^{136}$Xe.

The production yields for individual carbon spallation isotopes are well estimated with fits to the $\Delta T$ curves within $10^{3}$\,s for each isotope using the high statistics data in the large outer-LS volume. On the other hand, the same fit does not work for xenon spallation in the \mbox{Xe-LS} volume because there are many candidate long-lived isotopes whose individual yields are too small to be decomposed. To assess the total yields of all isotopes, we performed an MC simulation of muon-induced spallation using \texttt{FLUKA}~\cite{BOHLEN2014211, Ferrari:898301} and of the subsequent radioactive decays by \texttt{GEANT4}. The expected $\Delta T$ curve within $10^{6}$\,s and energy spectrum are calculated by adding the contributions of all produced isotopes as well as their daughters. The primary contributions are from $^{132}$I, $^{130}$I, $^{124}$I, $^{122}$I, $^{118}$Sb, $^{110}$In, and $^{88}$Y. Considering potentially large uncertainties in the MC-based total yields, we introduced a parameter to scale the long-lived spallation background rate ($\alpha_{\rm BG}$) in the fit to $0\nu\beta\beta$ decay discussed later. The systematic uncertainties on the relative yields are estimated from isotopic production cross sections for $^{136}$Xe spallation by protons at incident energies of 500\,MeV and 1\,GeV per nucleon~\cite{Giot2013,Napolitani2007}, giving an estimate of the energy spectral distortion from those errors. The MC study shows $(42.0 \pm 8.8)\%$ of long-lived spallation backgrounds are classified as LD, whereas only 8.6\% of uncorrelated events are mis-classified. This indicates that the LD analysis is useful for constraining $\alpha_{\rm BG}$.

\begin{figure}
\vspace{0.1cm}
\includegraphics[width=1.0\columnwidth]{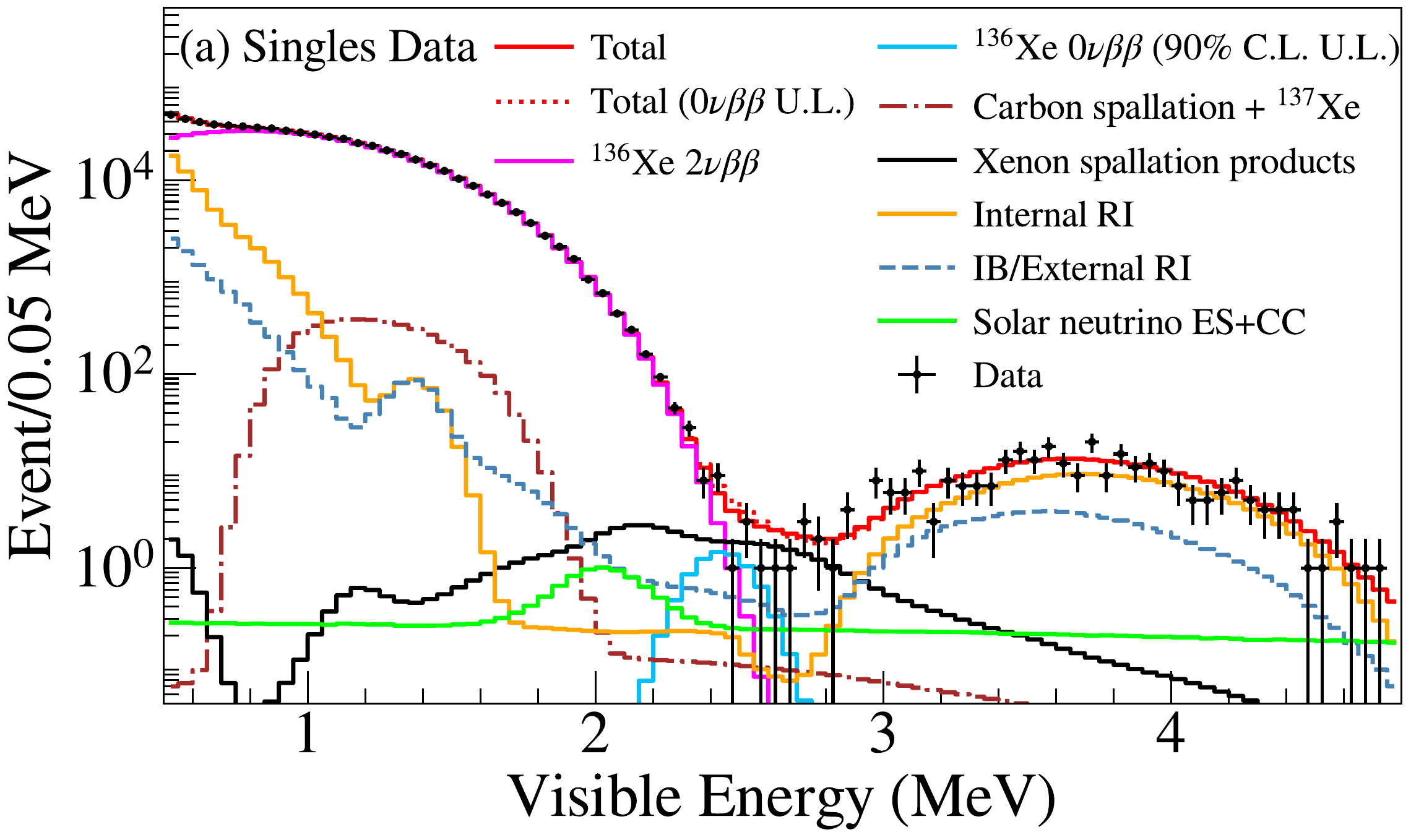}
\\
\vspace{0.3cm}
\includegraphics[width=1.0\columnwidth]{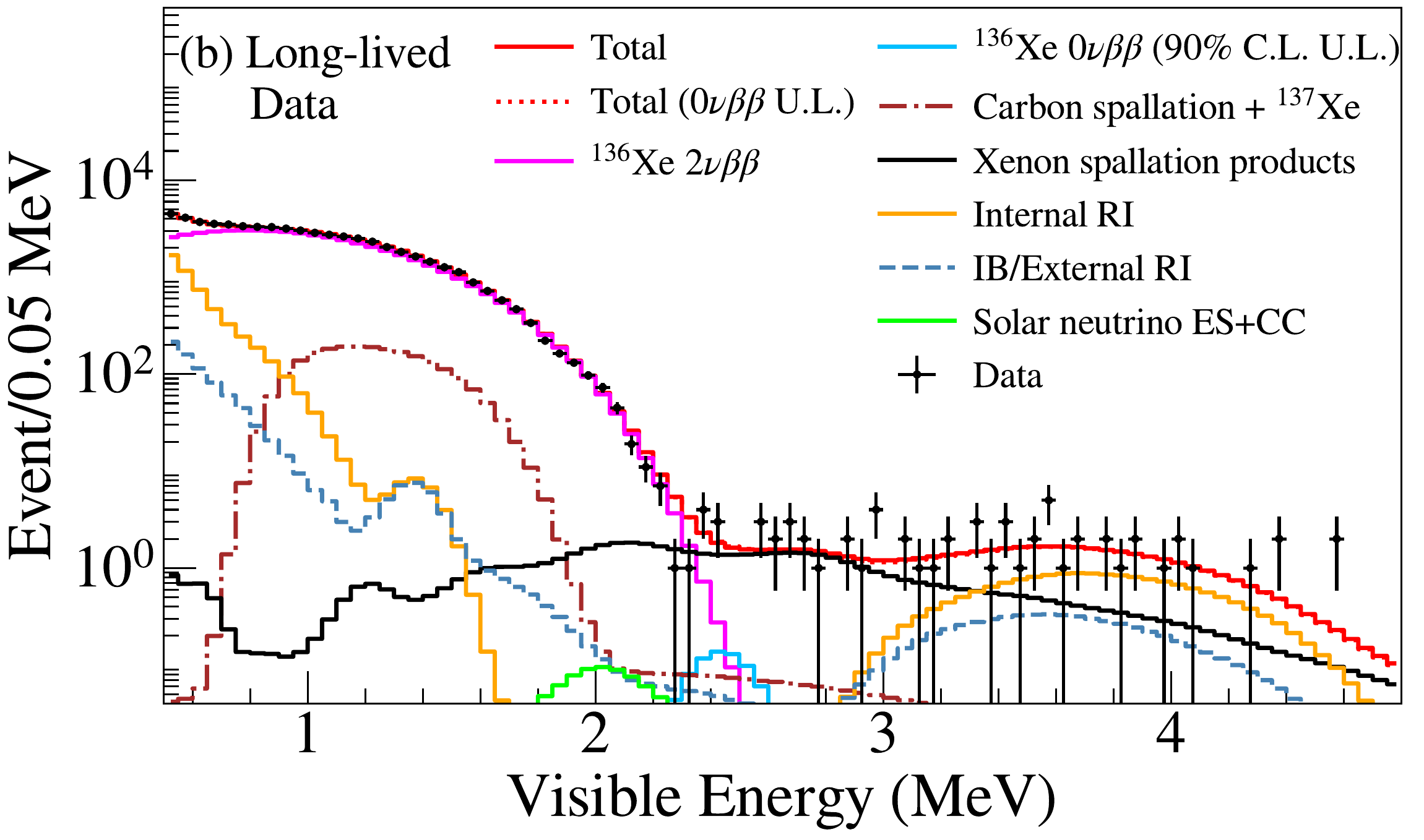}
\caption{Energy spectra of selected $\beta\beta$ candidates within a 1.57-m-radius spherical volume drawn together with best-fit backgrounds, the $2\nu\beta\beta$ decay spectrum, and the 90\% C.L. upper limit for $0\nu\beta\beta$ decay of (a) singles data (SD), and (b) long-lived data (LD). The LD exposure is about 10\% of the SD exposure.}
\label{figure:energy}
\end{figure}

\begin{table}[t]
\caption[]{Summary of the estimated and best-fit background contributions for the frequentist and Bayesian analyses in the energy region $2.35 < E < 2.70\,{\rm MeV}$ within the 1.57-m-radius spherical volume. In total, 24 events were observed.}
\label{table:background}
\begin{threeparttable}[h]
\begin{tabular}{lccc}
\hline
\hline
Background \hspace{1.0cm} & ~~~Estimated~~~ & \multicolumn{2}{c}{Best-fit} \\
 & & ~Frequentist~ & ~Bayesian~ \\
\hline
$^{136}$Xe $2\nu\beta\beta$ & - & $11.98$ & $11.95$ \\
\multicolumn{4}{c}{Residual radioactivity in \mbox{Xe-LS}} \\
\hline
$^{238}$U series & $0.14 \pm 0.04$ & $0.14$ & $0.09$ \\
$^{232}$Th series & - & $0.85$ & $0.87$ \\
\multicolumn{4}{c}{External (Radioactivity in IB)} \\
\hline
$^{238}$U series & - & $3.05$ & $3.46$ \\
$^{232}$Th series & - & $0.01$ & $0.01$ \\
\multicolumn{4}{c}{Neutrino interactions} \\
\hline
$^{8}$B solar $\nu$ $e^{-}$ ES & $1.65 \pm 0.04$ & $1.65$ & $1.65$ \\
\multicolumn{4}{c}{Spallation products} \\
\hline
Long-lived & $7.75 \pm 0.57$ \tnote{$\dagger$} & $12.52$ & $11.80$ \\
$^{10}$C & $0.00 \pm 0.05$ & $0.00$ & $0.00$ \\
$^{6}$He & $0.20 \pm 0.13$ & $0.22$ & $0.21$ \\
$^{137}$Xe & $0.33 \pm 0.28$ & $0.34$ & $0.34$ \\
\hline
\hline
\end{tabular}
\begin{tablenotes}
\item[$\dagger$] Estimation based on the spallation MC study.
This event rate constraint is not applied to the spectrum fit.
\end{tablenotes}
\end{threeparttable}
\end{table}

The $0\nu\beta\beta$ decay rate is estimated from a simultaneous likelihood fit to the binned energy spectra of SD and LD between 0.5 and 4.8\,MeV in hemispherical-shell volumes. The volumes are made by dividing the 2.5-m-radius fiducial volume into 20 equal-volume bins each in the upper and lower hemispheres. The contributions from major backgrounds in the \mbox{Xe-LS}, such as $^{85}$Kr, $^{40}$K, $^{210}$Bi, the $^{228}$Th-$^{208}$Pb subchain of the $^{232}$Th series, and long-lived spallation products, are free parameters and are left unconstrained in the spectral fit. The contributions from the $^{222}$Rn-$^{210}$Pb subchain of the $^{238}$U series and short-lived spallation products can vary but are constrained by their independent measurements. The parameters of the detector energy response model common to SD and LD are floated but are constrained by the $^{222}$Rn-induced $^{214}$Bi data. The energy spectral distortion parameter changing the relative contributions of the long-lived spallation isotopes is allowed to float freely. The uncertainty on the energy scale parameter is stringently constrained by the fit to the high statistics $2\nu\beta\beta$ events, so its impact on the background estimate is negligible.

To visualize the fit to the $0\nu\beta\beta$ signal and long-lived spallation backgrounds, the energy spectra of selected candidate SD and LD events within a 1.57-m-radius spherical volume (inner 10 equal-volume bins illustrated in Fig.~\ref{figure:vertex}(a)), together with the best-fit curves, are shown in Fig.~\ref{figure:energy}. The radial dependences of candidate SD events and best-fit background contributions in the $0\nu\beta\beta$ window are illustrated in Fig.~\ref{figure:vertex}(b). The exposure of $^{136}$Xe for SD in this volume is 0.510\,ton\,yr. The best-fit background contributions are summarized in Table~\ref{table:background}. We found no event excess over the background expectation. We obtained a 90\% confidence level (C.L.) upper limit on the number of $^{136}$Xe $0\nu\beta\beta$ decays of $<$\,$7.9$\,events ($<$\,$6.2$\,events in the range $2.35 < E < 2.70\,{\rm MeV}$), which corresponds to a limit of $<$\,$15.5$\,\PerTonYr~in units of $^{136}$Xe exposure, or $T_{1/2}^{0\nu\beta\beta} > 2.0 \times 10^{26}$\,yr (90\% C.L.). An analysis based on the Feldman-Cousins procedure~\cite{PhysRevD.57.3873_FCmethod} gives a slightly stronger limit of $2.3 \times 10^{26}$\,yr (90\% C.L.), indicating a limited impact of the physical boundary on the $0\nu\beta\beta$ rate in low statistics. An MC simulation of an ensemble of experiments assuming the best-fit background spectrum and including the high-background-period identification scheme indicates a median sensitivity of $1.3 \times 10^{26}$\,yr. The probability of obtaining a limit stronger than that reported here is 24\%. In addition to the frequentist analyses above, we also performed a statistical analysis within the Bayesian framework, assuming a flat prior for $1/T_{1/2}^{0\nu\beta\beta}$. The Bayesian limit and sensitivity are $2.1 \times 10^{26}$\,yr and $1.5 \times 10^{26}$\,yr (90\% C.L.), respectively.

We investigated the stability of the results by comparing the limits with different analysis conditions and background models. Alternatively, we also performed the analysis including the high-background period in the data with floated background contributions from $^{60}$Co and $^{214}$Bi. This data is separated into $\beta$-like and $\gamma$-like events, using particle identification provided by KamNet, and simultaneously fit to provide slightly improved half-life limits of $T_{1/2}^{0\nu\beta\beta} >2.7\times 10^{26}$\,yr and $T_{1/2}^{0\nu\beta\beta} >2.4\times 10^{26}$\,yr (90\% C.L.) for the background models with $^{60}$Co and $^{214}$Bi, respectively.

\begin{figure}[t]
\includegraphics[width=1.05\columnwidth]{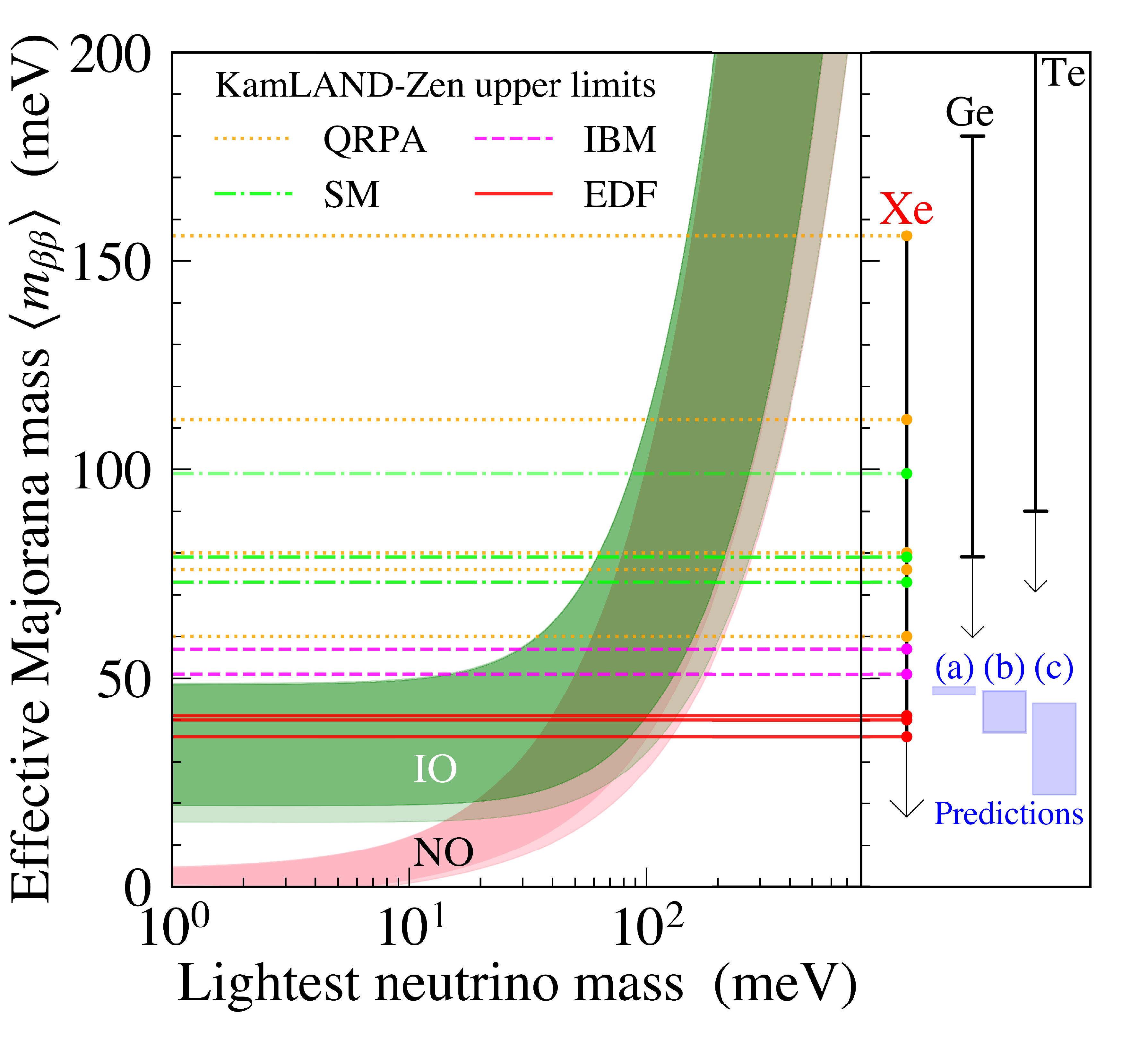}
\vspace{-0.7cm}
\caption{Effective Majorana neutrino mass $\left<m_{\beta\beta}\right>$ as a function of the lightest neutrino mass. The dark shaded regions are predictions based on best-fit values of neutrino oscillation parameters for the normal ordering (NO) and the inverted ordering (IO), and the light shaded regions indicate the $3\sigma$ ranges calculated from oscillation parameter uncertainties~\cite{DellOro2014,NuFIT2020}. The regions below the horizontal lines are allowed at 90\% C.L. with $^{136}$Xe from \mbox{KamLAND-Zen} (this work) considering an improved phase space factor calculation~\cite{Kotila2012,Stoica2013} and commonly used nuclear matrix element estimates: energy-density functional (EDF) theory~\cite{PhysRevLett.111.142501,PhysRevC.91.024316,Rodriguez2010} (solid lines), interacting boson model (IBM)~\cite{Deppisch2020,PhysRevC.91.034304} (dashed lines), shell model (SM)~\cite{PhysRevC.101.044315,Horoi2015,Menendez2009} (dot-dashed lines), and quasiparticle random-phase approximation (QRPA)~\cite{PhysRevC.102.044303,PhysRevC.91.024613,PhysRevC.87.045501,PhysRevC.87.064302,PhysRevC.97.045503} (dotted lines). The side panel shows the corresponding limits for $^{136}$Xe, $^{76}$Ge~\cite{Agostini2020}, and $^{130}$Te~\cite{Adams2022}, and theoretical model predictions on $\left<m_{\beta\beta}\right>$, (a) Ref.~\cite{Harigaya2012}, (b) Ref.~\cite{Asaka2020}, and (c) Ref.~\cite{Asai2020} (shaded boxes), in the IO region.}
\vspace{-0.3cm}
\label{figure:effective_mass}
\end{figure}

The combined fit of the KamLAND-Zen 400 and 800 datasets with the frequentist analyses gives a limit of $2.3 \times 10^{26}$\,yr (90\% C.L.) (see the Supplemental Material~\cite{SuppMat}). The best-fit scaling parameter for the long-lived spallation background rate is $\alpha_{\rm BG} = 1.35 \pm 0.23$, indicating good consistency between the MC-based prediction and the LD analysis. This combined analysis has a sensitivity of $1.5 \times 10^{26}$\,yr, and the probability of obtaining a stronger limit is $23$\%.  From the combined half-life limits, we obtain a 90\% C.L. upper limit of $\left<m_{\beta\beta}\right> < (36 \text{ -- } 156)\,{\rm meV}$ using the phase space factor calculation from~\cite{Kotila2012,Stoica2013} and commonly used nuclear matrix element estimates~\cite{PhysRevLett.111.142501,PhysRevC.91.024316,Rodriguez2010,Deppisch2020,PhysRevC.91.034304,PhysRevC.101.044315,Horoi2015,Menendez2009,PhysRevC.102.044303,PhysRevC.91.024613,PhysRevC.87.045501,PhysRevC.87.064302,PhysRevC.97.045503} assuming the axial coupling constant $g_{A} \simeq 1.27$. Figure~\ref{figure:effective_mass} illustrates the allowed range of $\left<m_{\beta\beta}\right>$ as a function of the lightest neutrino mass. For the first time, this search with $^{136}$Xe begins to test the IO band, and realizes the partial exclusion of several theoretical models~\cite{Harigaya2012,Asaka2020,Asai2020}, that estimate $\left<m_{\beta\beta}\right>$ based on predictions of the Majorana CP phases.

This Letter reported the first $^{136}$Xe $0\nu\beta\beta$ search, at 1\,ton\,yr exposure, in KamLAND-Zen 800 using almost double the amount of enriched xenon and a cleaner nylon balloon relative to KamLAND-Zen 400. Our improved sensitivity provides a limit that reaches below 50\,meV for the first time with certain nuclear matrix element calculations~\cite{PhysRevLett.111.142501,PhysRevC.91.024316,Rodriguez2010}, and is the first search for $0\nu\beta\beta$ in the inverted mass ordering region. Even though extensive efforts were made to analytically reject muon spallation, the sensitivity is limited mainly by long-lived spallation backgrounds. In the future, we plan to upgrade to dead-time-free electronics to detect muon-induced neutrons more effectively and enhance the rejection performance for xenon spallation backgrounds. Such improvements and continued observation in KamLAND-Zen will provide more stringent tests of the neutrino mass scale in the inverted mass ordering region.

\begin{acknowledgments}
The \mbox{KamLAND-Zen} experiment is supported by JSPS KAKENHI Grants No. 21000001, No. 26104002, and No. 19H05803; the U.S. National Science Foundation awards no. 2110720 and no. 2012964; the Heising-Simons Foundation; the Dutch Research Council (NWO); and under the U.S. Department of Energy (DOE) Grant No.\,DE-AC02-05CH11231, as well as other DOE and NSF grants to individual institutions. The Kamioka Mining and Smelting Company has provided service for activities in the mine. We acknowledge the support of NII for SINET4.
\end{acknowledgments}

\bibliography{DoubleBeta}

\end{document}